\documentclass[preprint,12pt]{aastex}

\shorttitle{Mira Planets}
\shortauthors{Struck, Cohanim, \& Willson}

\begin{document}


\title{Models of Planets and Brown Dwarfs in Mira Winds}


\author{Curtis Struck\altaffilmark{1}, Babak E. Cohanim, and Lee Anne
Willson\altaffilmark{2}}  
\affil{Department of Physics and Astronomy Department, Iowa State University,
Ames, IA 50011}


\altaffiltext{1}{Dept. of Physics and Astronomy,
  12 Physics Bldg., Iowa State Univ., Ames, IA 50011,
curt@iastate.edu} 
\altaffiltext{2}{lwillson@iastate.edu}


\begin{abstract}
We present numerical hydrodynamical models of the effects of planets
or brown dwarfs orbiting within the extended atmosphere and wind
formation zone of Mira variables. We find time-dependent wake dynamics
and episodic accretion phenomena which may give rise to observable
optical events and affect SiO maser emission.
\end{abstract}




\section{Introduction}

When stars like the Sun evolve to become AGB stars, their outer
envelopes swell to sizes of order an astronomical unit (au). During
the AGB phase the star becomes unstable to large amplitude radial
pulsations, and becomes a Mira or semi-regular variable (e.g.,
\citet{han94}).  These pulsations drive strong shock waves down the
steep pressure gradient outside the photosphere, producing an extended
atmosphere, and a dense, warm stellar wind (\citet{bow88}, also see
\citet{wil00}).  In the most luminous stars the wind driving is also
enhanced by radiation pressure on dust grains that form a few stellar
radii from the photosphere (e.g., \citet{bow88}, \citet{gai99},
\citet{wil00}).  During the Mira phase it is likely that the winds are
primarily radial since the massive, extended stellar atmosphere cannot
have significant rotation, and at this stage it is unlikely that a
globally ordered magnetic field strongly affects the dynamics of the
wind (see \citet{sok02}).

In recent years, about 80 extrasolar planetary systems have been
discovered \citep{but01}.  Generally, these systems have a single gas
giant planet, with a mass of up to $\simeq 15$ Jupiter masses ($M_J$),
and an orbital semi-major axis of commonly less than $1.0\  au,$ but
ranging up to $\simeq 4.0\  au$ (e.g., \citet{hat00}, \cite{fis02}).
Increasing numbers of brown dwarf stars have also been discovered
recently.  Most of these brown dwarfs are not companions of main
sequence Mira progenitors (but see \cite{fis02}, \cite{liu02}). The
more massive ``planets'' are in fact low mass brown dwarfs, and they
make up a significant fraction of the extrasolar 'planetary' systems.

The interaction between Mira winds and giant planets or low-mass brown
dwarfs with small orbital radii is an intriguing topic. This is
especially true in light of the distortion observed in the outflow in
the {\it{o Ceti}} system, apparently as a result of its (more distant)
white dwarf companion \citep{kar97}. The class of symbiotic systems
containing a Mira and a dwarf companion are also similar. Jura and
collaborators \citep[and refs. therein]{jur99} have studied other
interesting giant star binary systems, with molecular reservoirs and
dust rings.  We cannot hope to observe close planet systems in the
same detail as these related systems. As we will describe below,
observational signatures are likely to be indirect (see
\citet{wil01}).

Because planetary orbital periods (e.g., 3-30 yr.)  are roughly
similar to wind crossing times over radial scales of about 1.0 au (.5
- 3.0 yr.), massive planets will create substantial wakes, and perturb
the local wind flow. Within about 5 au, the motion of the extended
stellar atmosphere is still primarily oscillatory.  The oscillating
atmospheric gas elements and the propagating shocks interact with the
planetary bow shock and wake, so the resulting flow is complex and
very time-dependent. The planet accretes gas out of the stellar
wind, but the amount of mass accreted over the AGB lifetime is very
small compared to the planet's mass \citep[and below]{wil01}. Gas drag
on Jovian planets and brown dwarfs will also be small, unless they
orbit very close to the star, i.e., within the atmosphere
\citep{wil02}.

Nonetheless, the accretion may release enough energy, in a suitable
form, to give rise to observable events.  E.g., \citet{wil01} argue
that episodic optical flashes like those reported in the literature
(e.g., \citet{del98}, \citet{sch91}, \citet{ste01}) may be
produced. \citet{mou01} have also suggested that wake-like cometary
'exospheres' of giant planets, orbiting near their parent stars, might
be observable. The planetary perturbation may affect other still
observables, like the SiO maser emission that arises from the same
region \citep{str88}. To further understand both the observational
signatures and the basic dynamics of the interaction, we have
undertaken a program of hydrodynamic modeling of such systems, and
present the first results below.

Ultimately, the observational signatures of planets or brown dwarfs in
Mira winds could provide a new means of discovering such companions,
and of learning more about their fates in the late stages of stellar
evolution.

\section{Models}

The model results presented below were produced with a smoothed
particle hydrodynamics (SPH) code, in which hydrodynamic forces are
computed with a spline kernal on a grid with a fixed, uniform cell
size.  A standard artificial viscosity formulation is used to model
shocks.  The code is described in detail in \citet{str97}.

Both stellar and planet gravities were approximated as softened point
masses, though with small softening lengths. The SPH smoothing length
was set to 0.1 length units, where the unit of length was chosen to be
1.0 au.

Initially 38500 gas particles were placed over a quarter annulus in
the orbital plane of the planet, and on a circular coordinate grid
with constant radial intervals and a constant number of particles on
each azimuthal arc.  This distribution yields a $1/r^2$ density
profile (in three dimensions), similar to that observed.  The particles
were then given slight random offset from the initial grid and a small
random velocity in the plane, to smooth the distribution. Initial
radial and azimuthal velocities were zero except for this random part.

The models were run with a three dimensional code, but the model is
essentially two dimensional, since the particles were initialized in a
($z = 0$) plane, with zero vertical velocity. No vertical motions
developed throughout the runs.

An isothermal equation of state was used with the particle internal
energy set to a value corresponding to $T \simeq 1700 K$. In the
absence of dust the detailed Mira atmosphere models of \citet{bow88}
show that there are large temperature variations. However, if there is
even a small amount of dust, then the gas is roughly isothermal over
the radial range of the present model at about the adopted
temperature.

We neglect radiation pressure on grains, assuming that the size and
abundance of grains is small.  Bowen's models suggest that grains form
just inside the radial range considered here, and that radiation
pressure exceeds the outward push of pulsational shocks in the outer
part of this region in solar metallicity stars.  However, the
application of a nearly zero pressure upper boundary condition
suppresses the oscillatory motion and gives rise to a wind at the
largest radii in our model, with an overall flow much like Bowen's
models.

For the boundary conditions in these preliminary models we have
employed some simple approximations. First, the sides (near the x and
y axes) are reflective boundaries, which have little effect since the
motion is primarily radial. Secondly, the outer bound is initially at
a radius of 4.0 au, with the planet on a circular orbit of radius 3.5
au, and is moved steadily outward at a velocity of $V_{shock} = 2.0$
units (or about $14.\ km/s$) for the duration of the run. Because
pulsational shocks give particles near the outer bound outward
impulses on timescales shorter than the fractional free-fall time
needed to return to their initial positions, gas particles move
outward. The outer bound is moved outward to prevent excessive
interaction between it and the particles.

Thirdly, the inner boundary moves outward and inward on a cycloidal
trajectory from an initial radius of $2.0\ au$. The period of this
motion is close to that of a ballistic particle at that radius, so
particles in that region track it quite closely.

The final boundary condition is applied on a circle of radius $0.08
au$ around the planet.  Any particles moving within this circle are
assumed to be captured by the planet, and thus, are immediately
transferred to the planet center, and given the same velocity as the
planet.  The radius of this circle equals about 170 Jupiter radii, but
since it is slightly less than the adopted SPH smoothing length, it is
at the resolution limit of the simulation. Although we will continue
to refer to it as a planet, its mass was chosen to be $30 M_J$, which
is in the brown dwarf range. With this relatively large mass the
gravity of the planet is large enough to show all the essential
hydrodynamics on a grid of modest resolution which contains a
significant part of the orbit. We see the same hydrodynamic effects on
a smaller scale with less massive planets, but those calculations
require a smaller scale grid, or many more particles.



\section{Model Results}

\begin{figure}
\epsscale{0.68}
\plotone{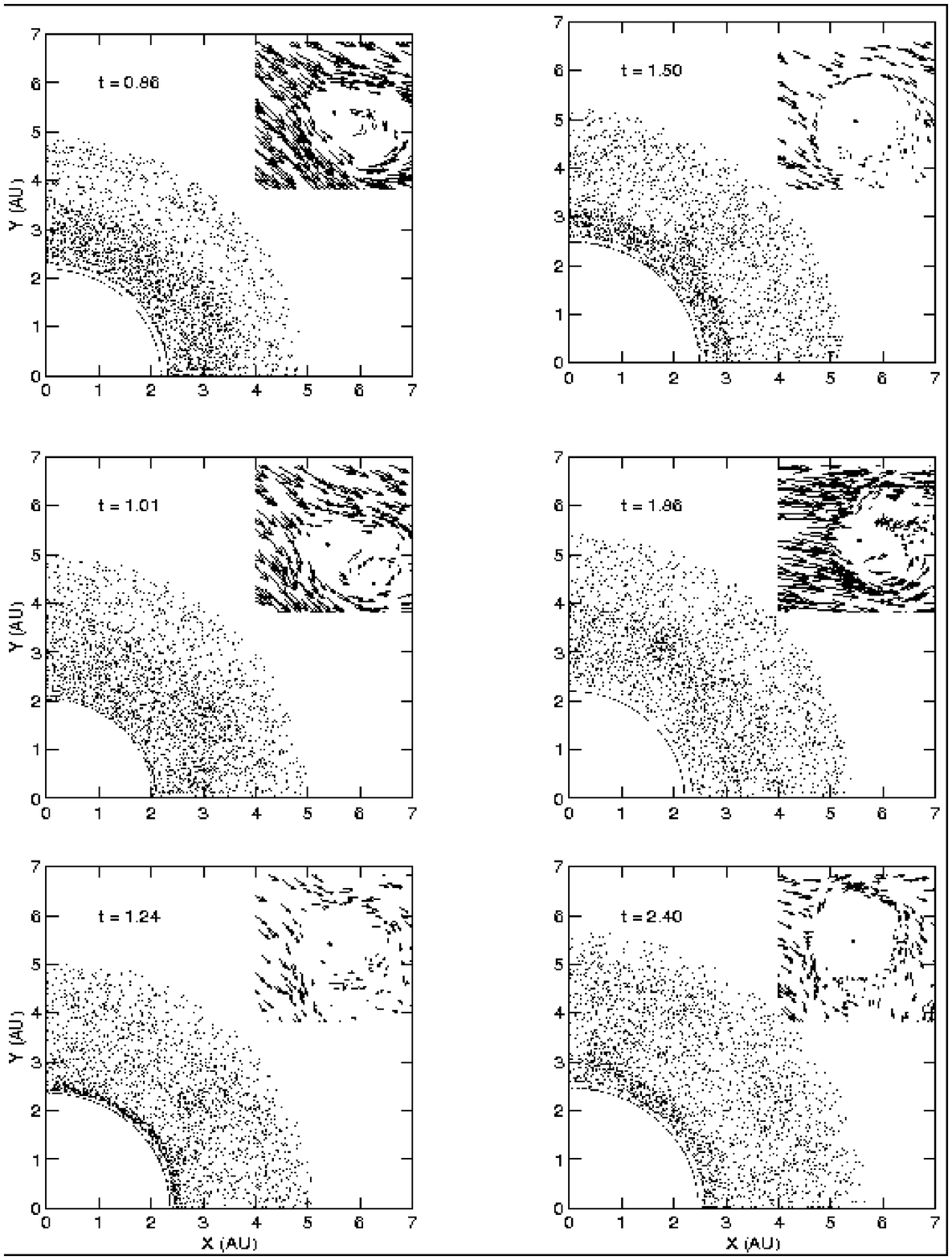}
\caption{Particle positions at 6 representative
times. Several pulsational shocks are shown, and the motion of the
planet and the evacuated region around it is also clear.  The
continually changing structure of the planetary wake is evident. The
accretion stream into the evacuted region also has a different
appearence in each snapshot.  This is emphasized in the insets, which
show selected particles in a small region around the planet, with
velocity vectors of length proportional to the particle's speed. Note
the strong counterclockwise inflow at t = 0.86 and 1.86 units, when
the shock has just passed the planet, and also the clockwise inflow at
t = 1.24 when the planet is in the middle of the post-shock
rarefaction. 
\label{fig1}}
\end{figure}

\begin{figure}
\plotone{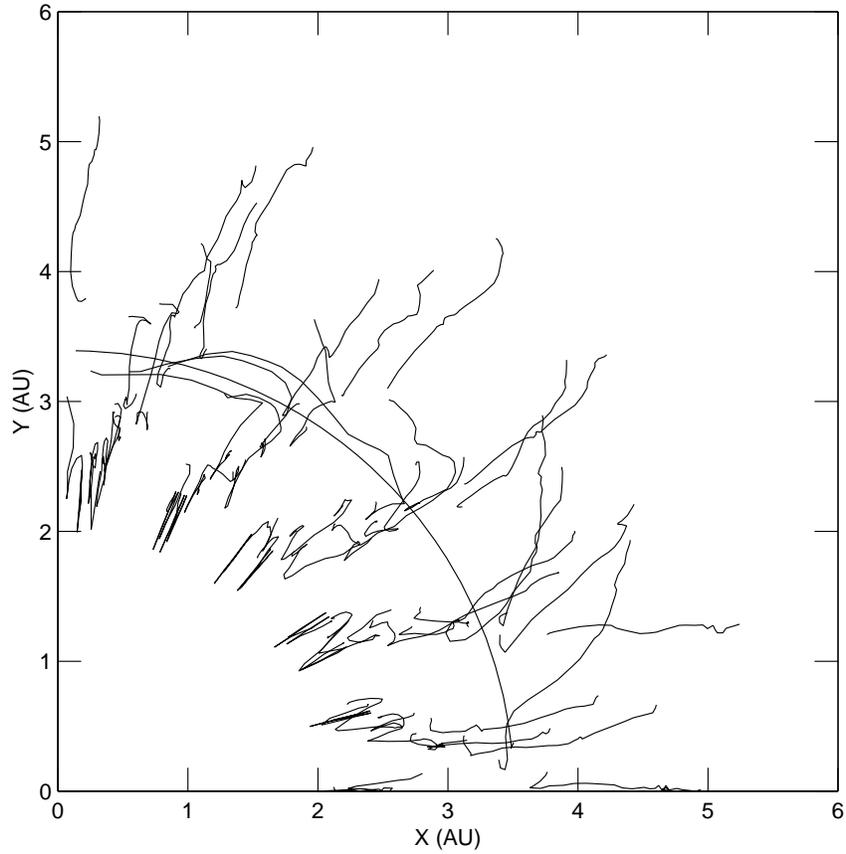}
\caption {The xy trajectories of a sample of a few tens of
particles. Note the oscillatory motion at smaller radii and the
generally outward motion at large radii.  This figure highlights the
effects of the planet's gravity on particles, which would have only
radial motions in the absence of the planet. The trajectory of the
planet is marked by a particle captured at early times.
\label{fig2}}
\end{figure}

Figure \ref{fig1} shows six snapshots from a representative model.
Over the course of this run the planet enters the grid from below and
moves around the quadrant, while several pulsational shocks are
generated at the base and move through the grid.  At the beginning of
the run (not shown) the planet enters the grid and begins to form a
bow shock and wake, while a shock is generated at the inner boundary.
This and subsequent pulsational shocks move through the grid and
fairly quickly set up a pattern of radial motions much like those in
Bowen's models \citep{bow88}.  That is, particle motions are primarily
oscillatory in the inner part of the grid and primarily outflowing in
the outer part(see Figure \ref{fig2}). Weak reflection shocks from the
outer boundary have a small effect on the radial motions of the
particles.

As an outward propagating shock sweeps over the planet the wake is
pushed into a more outward pointing configuration. As material falls
back in the rarefaction behind the shock, the wake reforms in an
azimuthal or slightly inward direction. At most times the wake has a
visible extent of at least $2\ au$.  Figure \ref{fig1} also shows that
downstream shock fronts are disturbed.

The insets in Figure \ref{fig1}, show how the planet accretes gas out
of the wake.  Wake accretion generally exceeds direct accretion.  In
most cases the accretion occurs via a single dominant stream.  This
stream has nonzero angular momentum relative to the planet, and so it
spirals into the planet, rather than falling radially.

The direction and amount of particle angular momentum relative to the
planet depends on the particle's radial velocity, which depends on the
phase of the local pulsation cycle.  In particular, as shown in the
Figure \ref{fig1} insets, infalling gas spirals around the planet
alternately in the clockwise and counter-clockwise directions.

\begin{figure}
\plotone{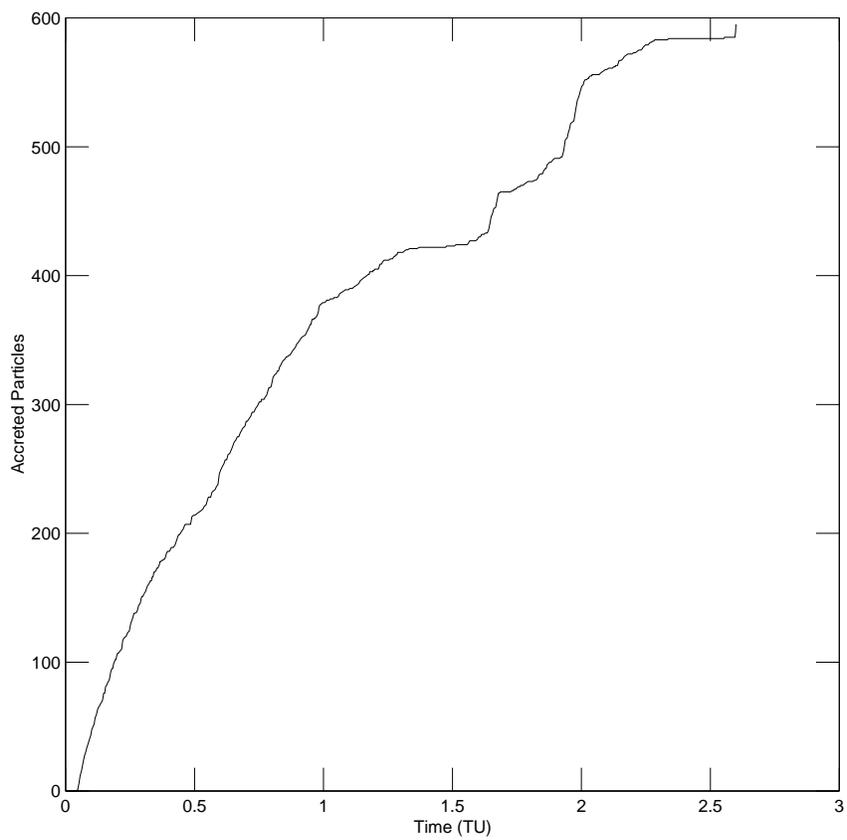}
\caption{Cumulative number of particles accreted by the
planet as a function of time.  Note the development of episodic
accretion. \label{fig3}}
\end{figure}

Figure \ref{fig3} shows the cumulative number of particles accreted as
a function of time.  Initially, while the wake is developing, the
accretion is quite steady. Later the accretion becomes episodic,
reflecting the time-dependence of the input stream from the wake. This
stream develops first in one sense, then as a result of the shocks and
radial flow reversals, the stream is first cut off, then
re-established in the opposite sense.

This alternating flow is reminiscent of the flip-flop instability
discussed in the literature on Bondi-Hoyle-Lyttleton accretion by
compact companions in OB star winds (see \citet{ben97}, \citet{ruf99}
and \citet{fog99} and references therein). The idea of reversing
accretion disks has also been considered in this context by
\citet{mur99}. However, in these cases, the flip-flop instability is
self-excited, and operates on an acoustic timescale. In the present
case, the 'flip-flop' is driven on a pulsational timescale.  It
appears that the peak to mean accretion rate ratio in the present case
is higher than in self-excited cases.

Figure \ref{fig2}, and other graphs not shown here, show that the
planet accelerates a significant amount of material in the azimuthal
direction.  Indeed, at the end of the run, wake particles can still be
detected at all azimuths by their azimuthal velocities. In the future
we plan to study this effect over a full orbit.

\section{Observational Consequences}

When the accretion stream flow changes direction, any accretion disk
that has formed around the planet from earlier streaming (not resolved
in our simulations), is likely to be severely disrupted. Most of
the accreted material probably falls onto the planet, dissipating its
infall energy.

We can estimate the rate of energy dissipation from the
simulation. Hydrodynamic variables like the density are dimensionless
in the code. We scale the initial density profile by assuming that it
is essentially the same as the time-averaged density profile in a
dust-free, model of \citet{bow88}.  In both cases the density profile
has approximately a $1/r^2$ form, and in the representative Bowen
model with $\dot{M} = 1.5 \times 10^{-7{\pm}2} M_{\odot} yr^{-1}$, the
density is about $2.0 \times 10^{-15}$ at 4.0 a.u.  We estimate the
total simulation volume as the initial area containing the particles
times a thickness of twice the SPH softening length (i.e., 0.2
au). Then with the given number of particles and the density profile
the mass per particle is $\simeq 2.5 \times 10^{21} g$.  We estimate
the energy released in an accretion burst as,

\begin{displaymath}
{\Delta}E = \frac{1}{2}{m_p}{N_b}{v_{ff}^2}
\simeq {6.8 \times 10^{37}}\   ergs
\end{displaymath}
\begin{equation} 
\times {\biggl(}\frac{m_p}{2.5 \times 10^{21} g}
{\biggr)} {\biggl(}\frac{N_b}{50}{\biggr)} {\biggl(}\frac{v_{ff}^2}
{330.\  km/s}{\biggr)}, 
\end{equation}
where $m_p$ is the particle mass, $N_b$ is the number of particles
accreted in the event, and $v_{ff}$ is the free-fall velocity at the
planet's surface.  The scaled value of $v_{ff}$ in the last equality
assumes a planet mass of $30 M_J$, and radius of $R_J \simeq 7.0
\times 10^9 cm$. The value of $N_b$ is that derived from the model,
with no correction for the cross section of the bow shock in three
dimensions, since the vertical extent assumed in making the particle
mass normalization is about the same as the bow shock radius.

The luminosity of the accretion burst is about,
\begin{displaymath}
L = \frac{{\Delta}E}{\tau_{ff}} \simeq {6.5 \times 10^{33}}  
\ ergs\  s^{-1}
\end{displaymath}
\begin{equation}
\times {\biggl(}\frac{3.0 hrs.}{\tau_{ff}}
{\biggr)} {\biggl(}\frac{M}{30.\ M_J}{\biggr)}^{3/2},
\end{equation}
where the free-fall time of 3.0 hours is evaluated at ten planetary
radii. Such short timescales cannot be resolved by the numerical
model. However, we have measured the angular momentum of gas elements
just before they are accreted. The circular orbital radii
corresponding to the measured angular momenta are of the order of
$3-10$ planetary radii, suggesting that this is the correct scale for
an accretion disk formed from this material.

The luminosity estimate above is about $1.7\ L_{\odot}$, yet there are
several reasons to believe that it could be higher.  In models with
dust, or larger amplitude pulsations, values of mass loss and wind
density can be increased by two orders of magnitude.  Moreover, with a
magnetotail to interfere with the downstream motion of particles in
the pulsational wave, the planet could acquire more. Thus, it is very
possible that the burst luminosity could substantially exceed a solar
luminosity in some cases. Any luminosity above $0.01 L_{\odot}$ is
potentially detectable in a Mira near minimum light.

Given the large free-fall velocity it is likely that much of the
energy is initially radiated in the UV and X-ray bands. A sphere of
radius $\simeq 10 R_J$ around the planet (or a hemisphere) will be
ionized. Much of this energy will be reprocessed and escape in the
form of optical continuum and emission line radiation.  Thus, the
simulation provides some support for the conjecture of \citet{wil01}
that short timescale optical flashes from Miras could be the result of
this process.

In reality, the accretion process must be very complex. For example,
the coupling of the partially ionized gas to the planetary magnetic
field could induce a variety of additional effects.  These might
include sporadic magneto-centrifugal jets, or energetic reconnection
events, but the relation of such processes to observables would
require more detailed models.

The complex dynamics induced by large planets almost certainly affect
another phenomenon of this region in Mira winds, the SiO maser
emission. Some aspects of this interaction were described by
\citet{str88}. For recent observational results see \citet{bob00},
\citet{dia01}, and references therein. Here we note that the nonlinear
perturbations in the planetary wake could have a significant effect on
the pumping of SiO molecules. The models show that density
perturbations persist and extend for a long distance behind the
planet, perhaps over most of the orbital circumference.  The density
amplitudes are also likely to be underestimated since the isothermal
models do not allow for the development of thermal instabilities
(e.g., mediated via SiO itself, see \citet{muc87}). Maser emission may
be enhanced near the planet in the accretion process, or indirectly,
from radiative pumping following accretion bursts.  As noted by
\citet{str88} the proximity of the planetary magnetosphere and
magnetotail, help account for high SiO maser polarizations.

Because of the nonlinear amplification in maser emission, it may
provide a particularly sensitive probe of Mira planets and close
stellar companions.  Monitoring of optical flashes, on the other hand,
provides a relatively simple means of searching many systems for
possible companions, especially brown dwarf companions. The discovery
and study of such companions would open a new window for the study of
the properties and population systematics of brown dwarfs and giant
planets.

\acknowledgments

We are grateful to G. H. Bowen for numerous helpful conversations and
unpublished model results. We are also grateful to Agnes Bishoff for
helpful conversations, and to the referee for pointing out the
flip-flop literature.

\clearpage



\end{document}